# LONG DISTANCE ATOMIC TELEPORTATION USING ENTANGLED COHERENT STATES AND CAVITY ASSISTED INTERACTION

Manoj K Mishra[1] and Hari Prakash [1,2]

[1] *Physics Department, University of Allahabad, India*

[2] *Indian Institute of Information Technology, Allahabad, INDIA.*

*Email: manoj.qit@gmail.com, prakash_hari123@rediffmail.com, hariprakash@iiita.ac.in*

**Abstract:** The schemes proposed by S. Bose *et al* [Phys. Rev. Lett. **83**, 5158 (1999)] and others for long distance atomic teleportation using cavity decay, gives message state dependent fidelity on successful attempt and in case of failure message state destroys. We propose a different scheme for teleportation of an atomic state from one cavity to a distant cavity by reflecting optical-pulse modes of an entangled coherent state (ECS) from single atom-cavity systems and detecting whether light is on or off. Unit success with unit fidelity is obtained for large mean photon number ($|\alpha|^2$). For small $|\alpha|^2$, however, there is some probability of failure but message state does not destroy, and this allows us to achieve unit success in a few repeated attempts. Unlike previous schemes, our scheme also enjoys advantages of deterministic generation of ECS, robustness of ECS against decoherence due to photon absorption, and it does not require many-stage cavity interactions and single photon detection ability.

## I. INTRODUCTION

Quantum teleportation, first due to Bennettt *et al* [1], is to transfer an unknown quantum state of a particle to another particle across space using long range EPR correlations [2] and a two-bit classical channel. Bowmeester *et al* [3], have experimentally demonstrated teleportation of a single photon polarized-state with the aid of standard bi-photonic Bell state (SBBS) that can be generated by parametric down conversion. In this experiment [3], authors used a linear beam splitter for the Bell state measurement (BSM) and reported success rate for teleportation equal to ½. The low success rate in such experiments where linear interaction are used for BSM is due to the fact that complete BSM of SBBS can not be performed with out nonlinear interaction, only two of the four Bell states can be discriminated and therefore success rate can not be larger than ½ [4]. Kim *et al* [5], reported 100% success rate using nonlinear sum frequency generation for BSM. The experiments [3, 5] with SBBS succeeded in proving the principle of quantum teleportation, but are commercially inapplicable due to low efficiency in production and detection of single photon, requirement of complex nonlinear interactions for complete BSM and sudden loss of entanglement due to photon absorption.

Another form of entangled states that attracted much attention are entangled coherent states (ECS) [6, 7] given by

$$|\psi_\pm\rangle_{a,b} = N_\pm[|\alpha,\alpha\rangle \pm |-\alpha,-\alpha\rangle]_{a,b}, \tag{1}$$

$$|\varphi_\pm\rangle_{a,b} = N_\pm[|\alpha,-\alpha\rangle \pm |-\alpha,\alpha\rangle]_{a,b}. \tag{2}$$

Normalization constants are given by

$$N_\pm = [2(1\pm x^4)]^{-1/2}, \tag{3}$$

where $x = \exp(-|\alpha|^2)$. Hirota *et al* [6], have shown that ECS are more robust against decoherence due to photon absorption than the SBBS. Another advantage of using ECS as quantum channel is that, by mixing two entangled modes of ECS over a 50:50 beam splitter followed by photon counting in output modes, one can precisely discriminate four ECS for large mean photon numbers of the order of $|\alpha|^2$ and, therefore, there is no need of nonlinear interaction for BSM if working with ECS of large mean photon numbers. Because of above mentioned advantages of ECS over SBBS, large numbers of schemes have been proposed to teleport single qubit superposed coherent state (SCS) [8] and two-qubit SCS [9] using ECS as quantum channel.

All these relevant works fall in optical domain, but from practical point of view, single-photon or SCS are not ideal for long term storage of quantum information as they are very difficult to keep in a certain place. On the other hand, it has been demonstrated [10] that a single atom can be trapped for 2-3 seconds inside an optical cavity. Thus, atoms are ideal for quantum information storage. For these reasons, number of schemes [11] for atomic teleportation using atom-cavity interactions and atoms as flying qubit have been proposed. Since atoms move slowly and interact strongly with their environment, these schemes are unable to perform long distance atomic teleportation and hence can not be used as link between two quantum processors working distant apart. However, S. Bose *et al* [12] have presented a novel scheme for teleporting quantum state of an atom trapped in an optical-cavity to second atom in another distant optical cavity.

This scheme involves mapping of atomic state to a cavity state with Alice, followed by the detection of photons leaking out from Alice's cavity and Bob's cavity (initially in maximally entangled atom-cavity state) by mixing over a beam splitter. The main shortcoming of this scheme is that the teleportation fidelity and success rate in this depends on the state to be teleported. Under reasonable cavity parameters and cavity decay time $t_D = 50\mu s$, success rate is near ½. Further Chimczak *et al* [13], pointed out the inefficiency of scheme [12] due to large damping values of currently available cavities that reduces the fidelity of state mapping from atom to cavity and discussed a modification using non-maximally entangled atom-cavity state with amplitudes chosen in such a way that compensates the damping factors due to state mapping. Although this resolves the effect of damping but gives very low success rate. In case of failure, in both schemes [12, 13] the message state is destroyed. Moreover, both schemes [12, 13] are expected to suffer decoherence due to photon absorption while propagating toward beam splitter. This requires many stages of atom-cavity interaction and single photon detection ability. For all these reasons, a *dream scheme for long distance atomic teleportation is required that (i) gives state independent teleportation fidelity and (ii) high success rate and (iii) conserves message state on failure thus permitting repeated attempts and (iv) does not need efficient single photon detection ability and (v) many matter-light interaction stages.* Along with these requirements, the scheme should use *(i) quantum channel that can be deterministically prepared and (ii) must be robust against photon absorption.* As discussed above since ECS are advantageous to be used as quantum channel than the SBBS and trapped atom in an optical cavity are ideal for quantum information storage, in this paper we propose a scheme for long distance atomic teleportation using ECS that fulfills most of the requirements mentioned above.

**II. TELEPORTATION SCHEME**

Before presenting our teleportation scheme, let us first review briefly the effect of an atom-cavity coupling over an external input coherent pulse [14]. Level configuration of atom trapped in a one-sided optical cavity is shown in Fig. 1, where $|g\rangle$ and $|f\rangle$ are the ground levels with different hyperfine spins and $|e\rangle$ is the exited level. The transition $|f\rangle \leftrightarrow |e\rangle$ is resonantly coupled to the cavity mode $a_c$, which is resonantly driven by an input coherent pulse $|\alpha\rangle$. The transition $|g\rangle \leftrightarrow |e\rangle$ is decoupled to the cavity mode $a_c$ due to large detuning from the hyperfine frequency. If initial joint state of atom and input pulse is $|g, \pm \alpha\rangle_{c,in}$, then input pulse is resonant with cavity and exact quantum optics calculation by D F Walls *et al* [15] shows that input pulse reflects with a phase $e^{i\pi}$. On the other hand, if initial state is $|f, \pm \alpha\rangle_{c,in}$, then due to strong atom cavity coupling, cavity mode $a_c$ is significantly detuned from the center frequency of the input pulse, thus input pulse reflects without any change in phase and pulse shape [14]. Mathematically these evolutions can be written as

$$|g, \pm \alpha\rangle_{c,in} \to |g, \mp \alpha\rangle_{c,out}, \quad |f, \pm \alpha\rangle_{c,in} \to |f, \pm \alpha\rangle_{c,out}. \tag{4}$$

Our detailed teleportation scheme is shown in Fig. 1. ECS $|\psi_+\rangle_{1,2}$ in mode 1 and 2 (Eq. 1) is produced by passing even coherent state $N_+(|\sqrt{2}\alpha\rangle + |-\sqrt{2}\alpha\rangle)_0$ in mode 0 through a 50:50 beam splitter BS1. Entangled mode 1 and 2 are sent to Alice and Bob respectively. Let us consider that Alice wishes to teleport message state of an atom in cavity C1 given by

$$|M\rangle_{C1} = [a|g\rangle + b|f\rangle]_{C1}, \tag{5}$$

with normalization condition,

$$|a|^2 + |b|^2 = 1, \tag{6}$$

to a second atom in a distant cavity C2, initially in state,

$$|+\rangle_{C2} = \tfrac{1}{\sqrt{2}}[|g\rangle + |f\rangle]_{C2}. \tag{7}$$

Joint state of Alice's and Bob's systems is given by

$$|\Phi\rangle_{1,2,C1,C2} = |\psi_+\rangle_{1,2}|M\rangle_{C1}|+\rangle_{C2}. \tag{8}$$

Bob reflects mode 2 from cavity C2 giving output mode 4, using Eq. (4) joint state can be written as

$$|\Phi\rangle_{1,2,C1,C2} = \tfrac{1}{\sqrt{2}}[|\varphi_+\rangle_{1,4}|g\rangle_{C2} + |\psi_+\rangle_{1,4}|f\rangle_{C2}]|M\rangle_{C1}. \tag{9}$$

Let us define four tripartite-entangled states composed of coherent field modes 1 and 4 on Alice's and Bob's side respectively and an atom in cavity C1 with Alice, as

$$\begin{aligned}
|\xi_\pm\rangle_{1,4,c1} &= \tfrac{1}{\sqrt{2}}(|\psi_+\rangle_{1,4}|g\rangle_{C1} \pm |\varphi_+\rangle_{1,4}|f\rangle_{C1}), \\
|\eta_\pm\rangle_{1,4,c1} &= \tfrac{1}{\sqrt{2}}(|\varphi_+\rangle_{1,4}|g\rangle_{C1} \pm |\psi_+\rangle_{1,4}|f\rangle_{C1}),
\end{aligned} \tag{10}$$

where $|\psi_+\rangle_{1,2}$ and $|\varphi_+\rangle_{1,2}$ are ECS given by Eq. (1, 2) respectively. Using Eq. (10) in Eq. (9) gives

$$|\Phi\rangle_{1,4,C1,C2} = \tfrac{1}{2}[|\eta_+\rangle_{1,4,C1}|M\rangle_{C2} + |\eta_-\rangle_{1,4,C1}(\sigma_Z|M\rangle_{C2}) \\ + |\xi_+\rangle_{1,4,C1}(\sigma_X|M\rangle_{C2}) + |\xi_-\rangle_{1,4,C1}(-i\sigma_Y|M\rangle_{C2})]. \tag{11}$$

Eq. (11) shows that, projective measurement in four tripartite-entangled states (Eq. 10) will give perfect teleportation.

To do this Alice reflects mode 1 from cavity C1 that gives output mode 3. Using Eq. (4) and representing Bob's atom in diagonal basis $|\pm\rangle = \tfrac{1}{\sqrt{2}}[|g\rangle \pm |f\rangle]$, the four tripartite-entangled states (Eq. 10) becomes

$$|\xi_\pm\rangle_{1,4,C1} \xrightarrow{\text{Reflecting mode 1 from cavity C1}} |\varphi_+\rangle_{3,4}|\pm\rangle_{C1}, \\ |\eta_\pm\rangle_{1,4,C1} \xrightarrow{\text{Reflecting mode 1 from cavity C1}} |\psi_+\rangle_{3,4}|\pm\rangle_{C1}. \tag{12}$$

Alice (Bob) then mixes mode 3 (4) with ancillary coherent state in mode 5 (6) over beam splitter BS2 (BS3) giving output modes 7 and 8 (9 and 10) as shown in Fig 1, after this Eq. (12) becomes

$$|\xi_\pm\rangle_{1,4,C1} \to [|\sqrt{2}\alpha,0,0,-\sqrt{2}\alpha\rangle + |0-\sqrt{2}\alpha,\sqrt{2}\alpha,0\rangle]_{7,8,9,10}|\pm\rangle_{C1}, \\ |\eta_\pm\rangle_{1,4,C1} \to [|\sqrt{2}\alpha,0,\sqrt{2}\alpha,0\rangle + |0-\sqrt{2}\alpha,0,-\sqrt{2}\alpha\rangle]_{7,8,9,10}|\pm\rangle_{C1}. \tag{13}$$

Output states on right side in Eq. (13) are orthogonal to each other for appreciable value of $|\alpha|^2$. Thus, four tripartite-entangled states defined in Eq. (10) can be discriminated by detecting that whether light is ON or OFF in modes 7, 8, 9 and 10 followed by an atomic measurement on cavity C1 in diagonal atomic basis $|\pm\rangle$. Thus quantum modes 1, 4 and C1 in Eq. (11) can be projected to any one of the four tripartite-entangled states defined in Eq. (6) just by reflecting mode 1 through cavity C1 followed by mixing of modes 3 and 4 with ancillary coherent state modes 5 and 6 respectively. Hence, we can achieve perfect teleportation for appreciable mean photon numbers. However, for small mean photon numbers ($|\alpha|^2$) due to shot-noise error [25], some times this measurement scheme fails.

To estimate success rate and resolve the problem of failure at small values of $|\alpha|^2$, we expand coherent state $|\pm\sqrt{2}\alpha\rangle$ into vacuum state $|0\rangle$ and state with nonzero numbers of photons ($|NZ_\pm\rangle$) given by

$$|\pm\sqrt{2}\alpha\rangle = x|0\rangle + \sqrt{1-x^2}|NZ_\pm\rangle. \tag{14}$$

Using Eq. (13) and Eq. (14) in Eq. (11), joint state of Alice's and Bob's system we get,

$$|\Phi\rangle_{7,8,9,10,C1,C2} = N_+[\{2x^2|0,0,0,0\rangle + x\sqrt{(1-x^2)}(|0,0,NZ_+,0\rangle + |NZ_+,0,0,0\rangle + |0,0,0,NZ_-\rangle$$
$$+ |0,NZ_-,0,0\rangle)\}_{7,8,9,10}|M\rangle_{C1}|+\rangle_{C2} + \tfrac{1}{2}(1-x^2)\{(|NZ_+,0,NZ_+,0\rangle + |0,NZ_-,0,NZ_-\rangle)_{7,8,9,10} \quad (15)$$
$$\times (|+\rangle_{C1}|M\rangle_{C2} + |-\rangle_{C1}(\sigma_Z|M\rangle_{C2}) + (|NZ_+,0,0,NZ_-\rangle + |0,NZ_-,NZ_+,0\rangle)_{7,8,9,10}$$
$$\times (|+\rangle_{c1}(\sigma_X|M\rangle_{C2}) + |-\rangle_{C1}(-i\sigma_Y|M\rangle_{C2})\}].$$

From Eq. (15), it is clear that two modes of the 7, 8, 9, and 10 are always in vacuum state. As mentioned above that teleportation requires detection of light that whether it is ON or OFF in field modes 7-10, and detection whether atom in cavity C1 is in $|+\rangle$ or $|-\rangle$. Thus measurement results can be classified into two groups,

Group I: Two field modes among 7-10 are detected as ON and atom in cavity C1 is detected in either of the states $|+\rangle$ or $|-\rangle$.

Group II: Three or all field modes among 7-10 are detected as OFF and atom in cavity C1 is detected in either of the states $|+\rangle$ or $|-\rangle$.

From Eq. (15), it is clear the when measurement results falls into group I, Bob's atom can be transformed to the original message state just by applying an appropriate unitary operation. Group I gives perfect teleportation with unit fidelity. We tabulated all possible measurement results corresponding to group I, Bob's atomic state after measurement and required unitary operator for prefect teleportation in table 1. The probability of successful teleportation ($P_s$) is given by summing the probability of occurrence of all measurement results corresponding to group I, and it is given by relation,

$$P_S = (1-x^2)^2(1+x^4)^{-1}. \quad (16)$$

However, for the measurement results corresponding to group II, teleportation fails. Probability of failure $P_f$, is given by

$$P_f = 2x^2(1+x^4)^{-1} = 1 - P_s. \quad (17)$$

But, from Eq. (15), it is clear that in such case before measurement on atom in cavity C1, the joint state of atoms in cavity C1 and C2 is given by $|M\rangle_{C1}|+\rangle_{C2}$. Thus message state of atoms in cavity C1 and initial state of the atom in cavity C2 remains conserved up to this stage. Keeping this in mind and to avoid failure, we now propose the following measurement scheme:

(*a*). Alice (Bob) detects field modes 7 & 8 (9 & 10) using detectors $D_7$ & $D_8$ ($D_9$ & $D_{10}$).

(*b*). Bob conveys his results to Alice using a two-bit classical channel.

(*c*). Alice looks over her measurement results and those conveyed by Bob, if any three or all modes among modes 7-10 gives OFF, she rejects the complete process. As already mentioned in such case initial message state and Bob's atomic state remain unchanged. Therefore, Alice does not make measurement on her atom in cavity C1 and starts new process with a fresh copy of ECS.

(*d*). However, if Alice finds two modes in nonzero photon states and remaining two in vacuum, she to measures her atom in cavity C1 and finally conveys two bit information to Bob about "atomic measurement result and the detector clicking result" through the same two-bit channel which was used earlier by Bob in step (*b*).

(*e*). Finally Bob performs the appropriate unitary operation on his atom in accordance of results conveyed by Alice and his own and generates exact replica of the information.

In this measurement scheme, step (*c*) avoids the failure by allowing us to repeat the complete process. In case if teleportation fails then by '*n*' number of attempts probability of success becomes

$$P_S^{(n)} = 1 - (P_f)^n. \tag{18}$$

We plotted $P_S^{(n)}$ with respect to $|\alpha|^2$ and '*n*'. Fig.2 shows that, in a single attempt '*n* =1' probability of success increases as $|\alpha|^2$ increases and becomes almost equal to unity for $|\alpha|^2 \geq 2.5$. This is due to the fact that for higher values $|\alpha|^2$ probability of having vacuum in coherent state becomes almost zero hence shot-noise error also becomes almost zero. However for small $|\alpha|^2$, probability of success is less then unity but increases rapidly with increasing number of attempts '*n*'. For example, at $|\alpha|^2 = 1$, in two attempts success rate rises to 0.8 and in three attempts it becomes almost unity. Thus unit success can be obtained in a single attempt for $|\alpha|^2 \geq 2.5$ or in finite number of attempts for low value of $|\alpha|^2 < 2.5$.

### III. EXPERIMENTAL FEASIBILITY AND DISCUSSION

Any ECS of the form, $N(|\alpha,\alpha\rangle + e^{i\varphi}|-\alpha,-\alpha\rangle)_{a,b}$, where $N = [2(1 + x^2 \cos\varphi)]^{-1/2}$, can, in principle be used in our scheme. This can be generated deterministically by illuminating a 50:50 beam splitter with SCS of the form, $N(|\beta\rangle + e^{i\varphi}|-\beta\rangle)_{a,b}$, where $\beta = \sqrt{2}\alpha$ with mean photon number of the order of $|\beta|^2$. It is well known that SCS can be generated from a coherent state by nonlinear interaction in a Kerr medium [16], but presently available Kerr nonlinearity is too small. For this reason schemes using very small Kerr nonlinearity [17] or without Kerr nonlinearity [14, 18-21] have been proposed. Alexi [18] and Nielsen [19] have experimentally demonstrated the generation of small SCS with $|\beta|^2 \approx 1$ by subtracting one photon from a squeezed vacuum. These kittens can be used to generate ECS with $|\alpha|^2 \geq 0.5$, with which at least three attempts are required to ensure 100% success of our scheme. These experimentally generated small SCS can be used for generating large SCS as described by Lund et al [20]. Recently, Alexi et al [21] proposed and experimentally demonstrated the generation of SCS with $|\beta|^2 \approx 10$ by photon number states, on the other hand Duan *et al* [14] numerically demonstrated that SCS with $|\beta|^2 \approx 5$ of high fidelity can be generated by reflecting optical pulses from a single atom-cavity. All these theoretical schemes [14, 16, 17, 20] and experiments [18, 19, 21] promises for the availability SCS at least with $|\beta|^2 \approx 5$ or ECS with $|\alpha|^2 \approx 2.5$, hence ensures almost 100% success of our scheme in single attempt.

Effect of atom-cavity coupling over an input optical pulse have been numerically simulated by Duan *et al* [14], including the influences of various experimental noise due to atomic spontaneous emission rate $\gamma_s$, photon pulse-shape distortion, cavity decay rate $\kappa$, and cavity mode-matching inefficiency. It is shown that for large pulse duration $T >> 1/\kappa$ the shape function of output pulse overlaps almost perfectly with the input pulse and is insensitive against random variation of coupling rate $g$ under typical variation range $3\kappa \leq g \leq 6\kappa$ (Fig. 2 in [14]). For reasonable atom-cavity coupling rate $g \approx 10\kappa$ and $\gamma_s = \kappa$ comparable with present technology, transformations in Eq. (4) are well approximated with fidelity ≥0.98 up to $|\alpha|^2 \approx 5$ and almost unity for $|\alpha|^2 \approx 2.5$ (Fig. 3 & 4 in [14]). This validates experimental feasibility of our scheme that gives perfect success rate for $|\alpha|^2 \approx 2.5$. Recent developments like fiber based Fabry-Perot cavity with $CO_2$ laser machined mirrors [22] can provide us efficient mode matching between the fiber and cavity modes, also existing method based on heterodyne detection of cylindrical transverse cavity modes allow us to optimize cavity mode matching to nearly 99.98% [23].

Finally, our scheme requires discrimination between vacuum state and coherent state i.e., detecting whether the light is ON or OFF. This can be done directly by photon counting using easily available Avalanche photon detectors (with shot-noise error for small $|\alpha|^2$). However, by using experimentally demonstrated [24] closed-loop measurement that combines photon counting with feedback-mediated optical displacements to discriminate vacuum and coherent states, we can reduce shot-noise error almost to zero for $|\alpha|^2 \geq 1$ (Fig 3 in [24]). Hence with this, our scheme can give almost 100% success rate in single attempt using ECS with $|\alpha|^2 \geq 2.5$ or SCS with $|\beta|^2 \geq 5$.

## IV. CONCLUSION

We presented an experimentally feasible scheme that, in principle, can teleport atomic state from one cavity to a distant cavity using ECS as quantum channel. For large $|\alpha|^2$ success rate is almost unity, for small $|\alpha|^2$ success rate becomes almost unity in a few attempts with unit fidelity. Unlike previous schemes based on cavity decay principal [12, 13], where (i) message state dependent fidelity with low success rate is obtained, or (ii) the message state is destroyed when teleportation fails and (iii) requires efficient single photon detection ability, our scheme gives message state independent unit fidelity and unit success rate in single attempt for $|\alpha|^2 \geq 2.5$. For low $|\alpha|^2$ there is nonzero probability of failure, but in such situation message state does not destroy and this leads to unit success within a few attempts. Also our scheme requires only threshold photon detectors that can discriminate between light 'ON' or 'OFF'.

**Acknowledgement**

We are grateful to Prof. N. Chandra and Prof. R. Prakash for their interest and stimulating discussions. Discussions with Ajay Kumar Yadav, Ajay Kumar Maurya and Vikram Verma are gratefully acknowledged. One of the authors (MKM) acknowledges the UGC for financial support under UGC-SRF fellowship scheme.

**Figure 1:**

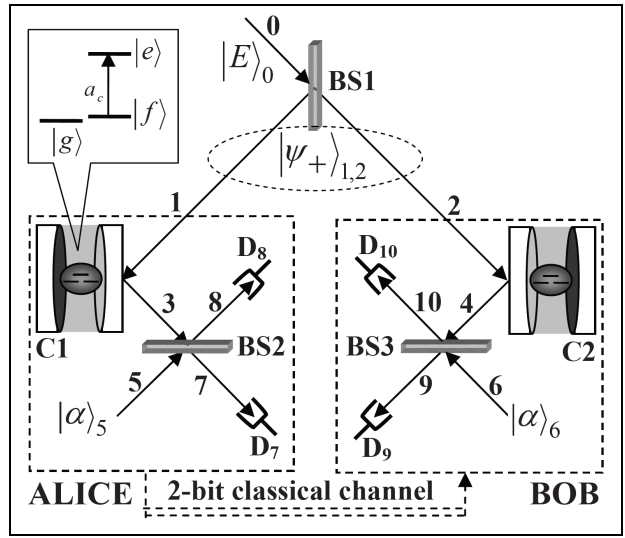

Fig.1: Scheme for teleportation of atomic-state trapped in cavity C1 to second atom in a distant cavity C2. Entangled coherent state ($|\psi_+\rangle_{1,2} = N_+[|\alpha,\alpha\rangle + |-\alpha,-\alpha\rangle]_{1,2}$) in modes 1 and 2 is produced by illuminating beam splitter BS1 with an even-coherent state ($|E\rangle_0 = N_+(|\sqrt{2}\alpha\rangle + |-\sqrt{2}\alpha\rangle)_0$) in mode 0. Numbers in bold represent the quantum mode. Inset shows level structure of atom. $D_{1,2,3,4}$ are photon detectors.

**Figure 2:**

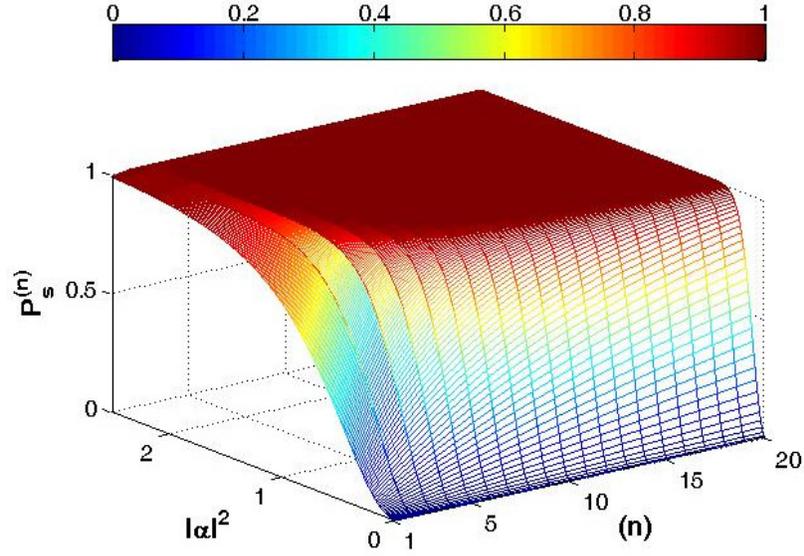

Fig.2: Shows variation of the success probability ($P_S^{(n)}$) for different numbers of attempts 'n' with $|\alpha|^2$.

**Table 1:**

Table1. Measurement results for successful teleportation. Tick stands for detection of nonzero photon (ON) and cross stands for detection of vacuum (OFF) by detectors. ± stands for atomic state in basis $|\pm\rangle$ and $\sigma$'s are Pauli matrices.

| Alice | | | Bob | | Bob's | Unitary |
|---|---|---|---|---|---|---|
| $D_7$ | $D_8$ | $D_\pm$ | $D_9$ | $D_{10}$ | Atomic state | operation |
| ✓ | ✗ | + | ✓ | ✗ | $M$ | $I$ |
| ✓ | ✗ | - | ✓ | ✗ | $\sigma_Z M$ | $\sigma_Z$ |
| ✗ | ✓ | + | ✗ | ✓ | $M$ | $I$ |
| ✗ | ✓ | - | ✗ | ✓ | $\sigma_Z M$ | $\sigma_Z$ |
| ✓ | ✗ | + | ✗ | ✓ | $\sigma_X M$ | $\sigma_X$ |
| ✓ | ✗ | - | ✗ | ✓ | $-i\sigma_Y M$ | $i\sigma_Y$ |
| ✗ | ✓ | + | ✓ | ✗ | $\sigma_X M$ | $\sigma_Z$ |
| ✗ | ✓ | - | ✓ | ✗ | $-i\sigma_Y M$ | $i\sigma_Y$ |